\documentclass[aps, superscriptaddress, prb, reprint]{revtex4-1}

\usepackage{graphicx}
\usepackage{dcolumn}
\usepackage{bm}

\begin{document}


\title{Nanoscale Magnetic Behavior Localization in Exchange Strength Modulated Ferromagnets} 



\author{B. J. Kirby}
\email{bkirby@nist.gov}
\affiliation{NIST Center for Neutron Research, National Institute of Standards and Technology, Gaithersburg, MD 20899, USA}

\author{L. Fallarino}
\affiliation{CIC nanoGUNE Consolider, E-20018 Donostia - San Sebastian, Spain}
\affiliation{Helmholtz-Zentrum Dresden-Rossendorf, Institute of Ion Beam Physics and Materials Research, Bautzner Landstrasse 400, 01328 Dresden, Germany}

\author{P. Riego}
\affiliation{CIC nanoGUNE Consolider, E-20018 Donostia - San Sebastian, Spain}
\affiliation{Departamento de F\`{i}sica de la Materia Condensada, Universidad del Pa\`{i}s Vasco, E-48080 Bilbao, Spain}

\author{B. B. Maranville}
\affiliation{NIST Center for Neutron Research, National Institute of Standards and Technology, Gaithersburg, MD 20899, USA}

\author{Casey W. Miller}
\affiliation{School of Chemistry and Materials Science, Rochester Institute of Technology, Rochester, NY, 14623}

\author{A. Berger}
\affiliation{CIC nanoGUNE, E-20018 Donostia - San Sebastian, Spain}


\date{\today}

\begin{abstract}
Although ferromagnetism is in general a long-range collective phenomenon, it is possible to induce local spatial variations of magnetic properties in ferromagnetic materials.  For example, systematic variation of the exchange coupling strength can be used to create systems that behave as if they are comprised of virtually independent segments that exhibit ``local" Curie temperatures.  Such localization of thermodynamic behavior leads to boundaries between strongly and weakly magnetized regions that can be controllably moved within the material with temperature.  The utility of this interesting functionality is largely dependent on the inherent spatial resolution of magnetic properties - specifically the distance over which the exchange strength and corresponding properties behave locally.  To test the degree to which this type of localization can be realized in materials, we have fabricated epitaxial films of Co$_{1-x}$Ru$_{x}$ alloy featuring a nanometer scale triangular wave-like concentration depth profile.  Continuous nanoscale modulation of the local Curie temperature was observed using polarized neutron reflectometry.  These results are consistent with mean-field simulations of spin systems that encompass the possibility of delocalized exchange coupling, and show that composition grading can be used to localize magnetic properties in films down to the nanometer level.   Since this is demonstrated here for an itinerant metal, we assert that for virtually any modulated magnetic material system, collective effects can be suppressed to length scales smaller than about 3 nm, so that magnetic behavior overall can be well described in terms of local material properties.
 \end{abstract}

\maketitle 
\section{Introduction} 
From a thermodynamic perspective, an array of ferromagnetically coupled spins can exhibit but a single magnetic transition temperature.  This stems from the fact that if the order parameter is nonzero somewhere, it is technically nonzero everywhere \cite{wang_onset_1992,skomski_curie_2000}.  However, from a practical point of view, the order parameter in one part of a material can of course be large while that in another region can become arbitrarily small \cite{ramos_new_1990, marcellini_influence_2009,le_graet_temperature_2015}.  Thus, a ferromagnet characterized by a Curie temperature ($T_{C}$) can exhibit a distribution of ``local" Curie temperatures $T_{C}$$^{\prime}$, corresponding to local variations in $J$, the exchange strength\cite{campillo_substrate_2001}.  In a ferromagnetic exchange graded system, $J$, and thereby $T_{C}$$^{\prime}$ vary continuously along a particular direction.  At temperatures ($T$) in between the minimum and maximum values of $T_{C}$$^{\prime}$,  the system exhibits a quasi phase boundary between ordered and disordered regions that moves reversibly along the gradient with varying $T$.  This case presents interesting functionality, as it allows for continuous control of the position of the magnetized - non-magnetized boundary in nanostructured materials.   Such behavior was recently realized experimentally in a compositionally graded NiCu alloy thin film \cite{kirby_spatial_2016}, while later work showed the device potential of a graded CoCr exchange well structure \cite{fallarino_magnetic_2017}. The prospective utility of exchange graded structures is strongly tied to the inherent spatial resolution - i.e. the degree of possible localization that is compatible with the collective nature of the ferromagnetic state. Experiments in Ref.~\onlinecite{kirby_spatial_2016} were conducted for a sample with a linear gradient spanning 100 nm, but were found to be consistent with nearest-neighbor mean-field simulations suggesting that relevant magnetic non-uniformities were localized down to a length scale of a few nm.   Such a high degree of localization is counterintuitive, as the specific system studied was metallic, with delocalized spin-polarized states.  

This curious localization has motivated us to explore the evolving thermodynamically ordered states that occur in metallic ferromagnets with composition gradients carefully controlled down to the nm length scale.  We have fabricated Co$_{1-x}$Ru$_{x}$ films with $x$ modulated in a triangular waveform along the growth direction.  For samples with modulation length scales down to 4.9 nm, neutron scattering measurements confirm the precise nanostructuring, and reveal continuously modulated magnetization profiles.  The minima and maxima of these profiles exhibit distinct local Curie temperatures, demonstrating localization down to approximately 3 nm.

\section{Sample Preparation}
\begin{figure}
\includegraphics[scale=1.0]{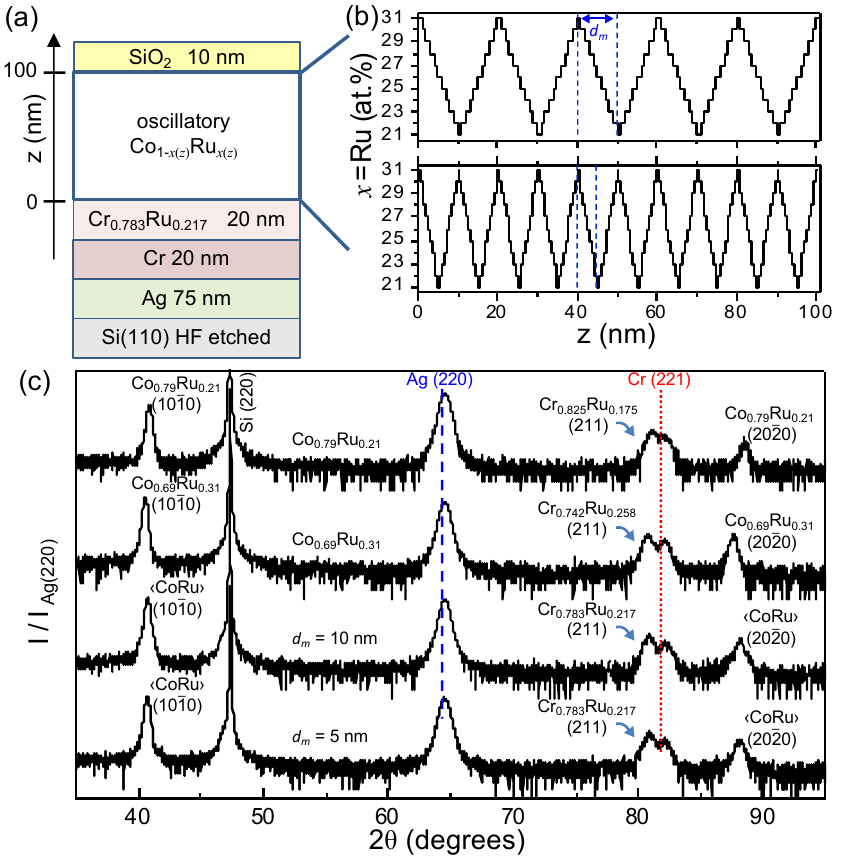}
\caption{\label{xrd}  (a) Layer structure of the samples.  (b)  Ru modulation for the $d_{m}$ = 10 nm (top) and $d_{m}$ = 5 nm (bottom) samples.  (c)  X-ray diffraction spectra for (from top) uniform $x$ = 0.21, uniform $x$ = 0.31, modulated $d_{m}$ = 10 nm and modulated $d_{m}$ = 5 nm films.  $\langle$CoRu$\rangle$ refers to peaks associated with the average Co$_{1-x(z)}$Ru$_{x(z)}$ structure.
} 
\end{figure}

We have previously shown that composition grading of CoRu can be used to create multilayer samples with nanoscale modulation of the saturation magnetization \cite{fallarino_modulation_2018}.  In this work, we consider the temperature dependent magnetic properties of this type of sample and the corresponding magnetic depth profile evolution. Moreover, we provide a detailed investigation of the lower limit of magnetic properties localization in such a composition graded ferromagnet.  CoRu alloy is ideal for this study, as it is a very simple ferromagnet with easily tunable magnetic properties \cite{pierron-bohnes_observation_1997,hashizume_observation_2006,qadri_structural_2007,eyrich_exchange_2012,idigoras_magnetic_2013}.   Over a wide range of $x$, Co$_{1-x}$Ru$_{x}$ forms a stable solid solution with the hcp crystal structure characteristic of pure Co, with both $T_{C}$ and saturation magnetization $M_{S}$ that decrease almost linearly with $x$.  Further, Co$_{1-x}$Ru$_{x}$ can be grown with the hcp crystal structure and (10$\bar{1}$0) orientation such that there is a single in-plane easy axis.  This makes the magnetostatic energy essentially irrelevant, leading to magnetic behavior well described by a macrospin model \cite{idigoras_collapse_2011,berger_transient_2013,riego_metamagnetic_2017}.  Epitaxial thin film samples were grown onto Si (110) oriented substrates by means of room temperature sputter deposition under a pressure of 0.4 Pa of pure Ar.  The layer structure of the system is shown in Figure~\ref{xrd}(a).   
Underlayers of Ag and Cr were deposited on the substrates to promote highly oriented (211) growth of a Cr$_{0.783}$Ru$_{0.217}$ which in turn served as a template for epitaxial growth of 100 nm of a (10$\bar{1}$0) Co$_{1-x(z)}$Ru$_{x(z)}$ compositionally modulated layer.  The samples were capped with a 10 nm protective SiO$_{2}$ layer.  Modulation of $x$ was achieved through power variation of Ru during co-sputtering of the Co and Ru, with the average Ru concentration 1.2 times that of the underlying Cr$_{1-y}$Ru$_{y}$ layer, the ideal ratio for epitaxial growth \cite{idigoras_magnetic_2013}.  The Co$_{1-x(z)}$Ru$_{x(z)}$ modulation scheme is depicted in Fig.~\ref{xrd}(b).  The Ru concentration varies periodically from $x$ = 0.21 to $x$ = 0.31 with a triangular waveform, as confirmed with neutron scattering (see below). The defining characteristic of the samples is the modulation distance $d_{m}$ between minima and maxima in $x$ (i.e. half the wavelength).  For this work, we consider two samples with nominal modulation distances $d_{m}$ = 10 nm (Fig.~\ref{xrd}(b) top) and $d_{m}$ = 5 nm (Fig.~\ref{xrd}(b) bottom), as well as uniform $x$ = 0.21 and $x$ = 0.31 reference samples grown using an identical underlayer sequence.

Cu K$_{\alpha}$ x-ray diffraction measurements confirm the epitaxial growth quality, as shown in Fig. \ref{xrd}(c). Well-defined peaks are observed indexed to Si (220), Ag (220), Cr (211), Cr$_{1-y}$Ru$_{y}$ (211), Co$_{1-x(z)}$Ru$_{x(z)}$ (10$\bar{1}$0), and (20$\bar{2}$0) crystal planes. Despite the complex depth-dependent structure, both the presence of second order Co$_{1-x(z)}$Ru$_{x(z)}$ peaks, and the absence of peaks corresponding to non-epitaxial crystal orientations demonstrate excellent crystallographic order.  
  
\begin{figure}
\includegraphics{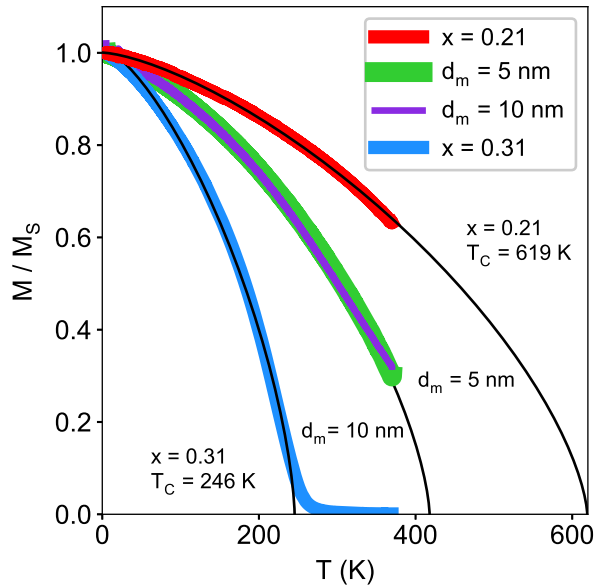}
\caption{\label{SQUIDVSM_MvT}  Temperature-dependent magnetizations of the modulated and control samples, as measured with SQUID in 5 mT.  Thin black lines are fits to Eq.~\ref{Kuzmin}, used to estimate $T_{C}$. }
\end{figure}

The temperature-dependent easy-axis magnetizations $M$($T$) for these samples were measured in a 5 mT in-plane field using a superconducting interference device magnetometer (SQUID), and are shown in Figure~\ref{SQUIDVSM_MvT}.   
The magnetizations are normalized to the low temperature values to more clearly depict differences in temperature dependence.  While measurements were restricted to $T$ $<$ 370 K to avoid sample damage, clear trends are observed.   The uniform $x$ = 0.21 reference sample exhibits a much larger $T_{C}$ than does the uniform $x$ = 0.31 sample, with the modulated $d_{m}$ = 10 nm and $d_{m}$ = 5 nm curves falling in between.  Values of $T_{C}$ were estimated quantitatively using  the method of Kuz'min \cite{kuzmin_shape_2005},
\begin{equation}
\frac{M}{M_{S}}=\left[1 - s\left( \frac{T}{T_{C}}\right)^{1.5} - (1-s)\left( \frac{T}{T_{C}} \right)^{p}\right]^{0.33}.
\label{Kuzmin}
\end{equation}
The shape parameters $s$ = 2.9 and $p$ = 1.9 were determined from the best-fit to the uniform $x$ = 0.31 curve (the only one with $T_{C}$ in our accessible temperature window), and were subsequently left fixed for fits of the other $M$($T$) curves (including those determined from PNR, as discussed in Section V).  As such, we make the assumption that $M$($T$) for Co$_{1-x}$Ru$_{x}$ alloys of similar but different $x$ should fall into the same universality class and exhibit the same critical exponent - i.e. all of the magnetization curves should exhibit the same shape \footnote{Details of the magnetometry measurements can be found in Section I of the online Supplemental Material}.  The best fits using Eq. 1 are shown as solid black lines, and indicate $T_{C}$ = 230 K for $x$ = 0.31 and $T_{C}$ = 560 K for $x$ = 0.21, generally consistent with reference \onlinecite{idigoras_magnetic_2013}.  For the modulated samples, such $T_{C}$ estimation is less meaningful, as we expect the net $M$($T$) to correspond to a superposition of distinct curves.  However, for the temperature range covered, the modulated samples exhibit very similar, smoothly varying $M$($T$) curves, that are very similar to what would be expected from a random alloy of the same average composition.

\section{Polarized Neutron Reflectometry}

Since the magnetic modulation of the $d_{m}$ = 10 nm and $d_{m}$ = 5 nm samples cannot be confirmed with conventional magnetometry alone, we characterized the magnetic and structural depth profiles using polarized neutron reflectometry (PNR).  Measurements were performed on the PBR beamline at the NIST Center for Neutron Research.  Samples were mounted in a cryostat in the presence of an in-plane saturating magnetic field $\mu_{0}H$ = 0.5 T aligned along the easy axis, and scans were carried out over a temperature range of 50 K - 300 K.  Using an Fe/Si supermirror / Al-coil assembly, a monochromatic 0.475 nm neutron beam was spin polarized parallel (+) or antiparallel (-) with respect to $H$ and was specularly reflected from the sample surface.  The reflected beam was spin analyzed (+ or -) using a second supermirror / coil assembly, and detected using a $^{3}$He tube.  The non spin-flip reflectivities $R^{++}$ and $R^{--}$ were measured as functions of wavevector transfer $Q$ along the sample surface normal. Data shown in this work was corrected for background, beam footprint, and beam polarization efficiency.  $R^{++}$($Q$) and $R^{--}$($Q$) can be calculated exactly from their respective depth ($z$) dependent scattering length densities \cite{majkrzak_polarized_2006}, which have nuclear and magnetic components, 
\begin{equation}
\rho^{(\pm)} = \rho_{nuc} \pm \rho_{mag}.
\label{rho}
\end{equation}
The nuclear scattering length density is indicative of the nuclear composition, and is defined
\begin{equation}
\rho_{nuc} = \sum_{i} n_{i}b_{i},
\label{rhoN}
\end{equation}
where $n$ is the number density, $b$ is the isotope specific nuclear scattering length, and the summation is over each type of isotope present in the material.  Values of $b$ are known for all isotopes discussed here \cite{fuess_neutron_2006}.   Thus, by assuming known bulk values of $n$, we can calculate expected values of $\rho_{nuc}$ for each material in our samples. The magnetic scattering length density is proportional to the in-plane component of the sample magnetization ($M$) parallel to the neutron spin axis (i.e. parallel to $H$),
\begin{equation}
\rho_{mag} = C M,
\label{rhoM}
\end{equation}
where $C$ = 2.91 $\times$ 10$^{-7}$ for $\rho$ in units of nm$^{-2}$ and $M$ in kA m$^{-1}$.  Thus, model fitting of the specular reflectivity can be used to determine the compositional and magnetic depth profiles.  For this work, fitting was performed using the Refl1D software suite \cite{kirby_phase-sensitive_2012}, with parameter uncertainty determined using a Markov chain Monte Carlo algorithm \cite{vrugt_accelerating_2009}.  All scattering data is shown with error bars corresponding to 1 standard deviation.  All reported uncertainties associated with fitting parameters correspond to 2 standard deviations.

\section{Bragg Peak Analysis}
\begin{figure}
\includegraphics[scale=1,trim = .45cm .5cm 0cm .25cm,clip ]{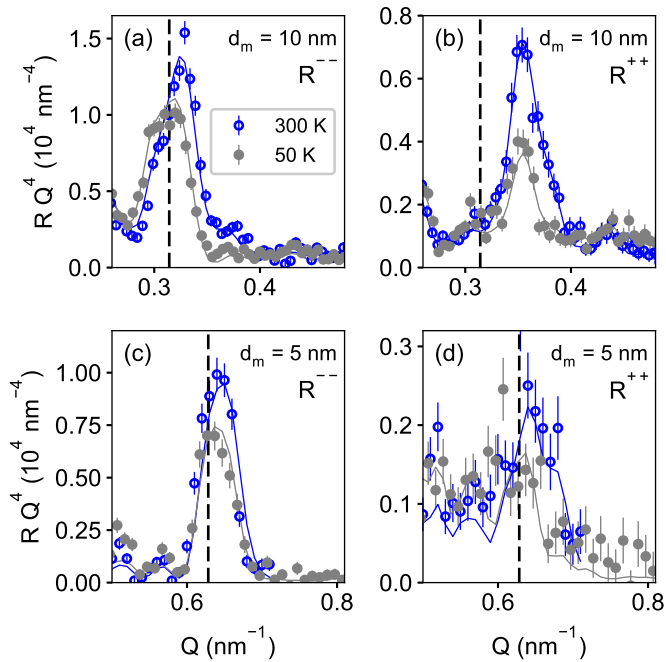}
\caption{\label{braggpeaks}First order Bragg peaks for the $d_{m}$ = 10 nm (a-b) and $d_{m}$ = 5 nm samples (c-d).  Solid lines are fits to the data corresponding to dynamical modeling of the full reflectivities.  Dashed vertical lines indicate the expected position of the first order Bragg peaks if dyanmical effects are neglected.}
\end{figure}

Before discussing quantitative modeling of the PNR data, we consider a simple qualitative analysis of the multilayer Bragg scattering. Neglecting dynamical effects (i.e. effects due to proximity to the reflectivity critical edge\cite{zabel_x-ray_1994,treimer_wavelength_1996}), first order Bragg scattering associated with the modulated structure shown in Fig. \ref{xrd}(b) would be expected at $Q_{m}$ = $\frac{\pi}{d_{m}}$.  Figure~\ref{braggpeaks} shows the observed scattering at 300 K and 50 K near $Q_{m}$ (depicted as dashed vertical lines) for each sample and spin state.  
The data are shown multiplied by $Q^{4}$ to compensate for the Fresnel decay of the reflectivity and better visualize the scattering across an extended $Q$-range.  Bragg peaks are indeed observed near $Q_{m}$, but all are somewhat positively $Q$-shifted.   As shown below with exact modeling of the reflectivity (corresponding to the solid lines in Fig.~\ref{braggpeaks}-\ref{RvQ}), this shifting is due both to $d_{m}$ values that are slightly smaller than the nominal deposition values, and the aforementioned dynamical effects. For both samples, and for both spin states, the Bragg peak intensity ($I^{\pm}$) increases significantly with increasing $T$,  
\begin{equation}
I^{\pm}(300 \: \textrm{K}) > I^{\pm}(50 \: \textrm{K}).
\label{TdepI}
\end{equation}
This alone can be used to infer that $T_{C}$$^{\prime}$ must be depth-dependent.   If we imagine that the triangle waveform grown into the Co$_{1-x(z)}$Ru$_{x(z)}$ layers is smoothed by some amount of interlayer roughness, the resulting profile should approximate a sinwave.  A sinusoidally modulated superlattice exhibits only a first order Bragg peak \cite{lynn_irongermanium_1976}, with intensity 
\begin{equation}
I^{\pm} \: \propto \: \frac{n \, d_{m}}{2}\left[\rho_{max} - \rho_{min}\right]^{2}.
\label{sinwave}
\end{equation}
For our system, we define``max" as the $x$ = 0.21 region (as in exhibiting the maximum magnetization), and``min" as the $x$ = 0.31 region.  The quantity we are most interested in is the magnetization modulation, 
\begin{equation}
\Delta_{mag} = M^{max} - M^{min}.
\label{deltaM}
\end{equation}
Similarly, we can define the nuclear modulation
\begin{equation}
\Delta_{nuc} = \rho_{nuc}^{max} - \rho_{nuc}^{min}.
\label{deltaN}
\end{equation}
Using Eq.~\ref{rho}, Eq.~\ref{sinwave} can then be rewritten in terms of the nuclear and magnetic modulation,
\begin{equation}
I^{\pm} \: \propto \: \frac{n \, d_{m}}{2}\left[\Delta_{nuc} \pm C\Delta_{mag}\right]^{2}.
\label{polsinwave}
\end{equation}
Substituting (\ref{polsinwave}) into (\ref{TdepI}), and taking into account that the nuclear composition is $T$-independent and that the region with more Co must exhibit a higher magnetization at all $T$ reveals that the magnetic modulation must be larger at high $T$,
\begin{equation}
\Delta_{mag}(300 \: \textrm{K}) >  \Delta_{mag}(50 \: \textrm{K}).
\label{Mineq}
\end{equation}
To relate $T_{C}$ and $M$, we assume that the minima and maxima are ferromagnetic, with monotonically decreasing $M$($T$) (e.g. as described by Eq. \ref{Kuzmin}) characterized by $x$-independent critical exponents.  In this case, $T_{C}$ is merely a scaling factor, and substitution of Eq \ref{Kuzmin} into Eq. \ref{Mineq} implies that,
\begin{equation}
T_{C} \,^{\prime \, max} > T_{C}\, ^{\prime \, min}.
\label{Tcmod}
\end{equation}
Thus, even without detailed model fitting, we can directly show that $T_{C}$$^{\prime}$ is modulated at the nanoscale, at least down to distances of approximately 5 nm.

\section{Reflectivity Model Fitting}
\begin{figure}
\includegraphics[scale=1]{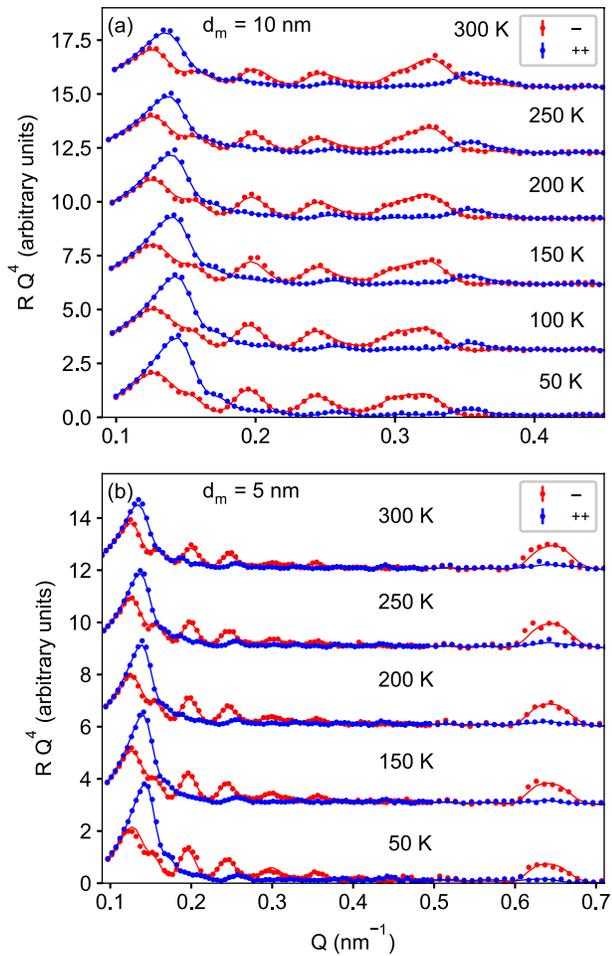}
\caption{\label{RvQ}Temperature-dependent reflectivities measured in 0.5~T for the (a) $d_{m}$ = 10 nm and (b) $d_{m}$ = 5 nm samples.  Fits correspond to the depth profiles in Fig.~\ref{profiles}, and are represented as solid lines through the data.  Reflectivities and fits are multiplied by $Q^{4}$, and are vertically offset for clarity.}
\end{figure}

\begin{figure}
\includegraphics[scale=1]{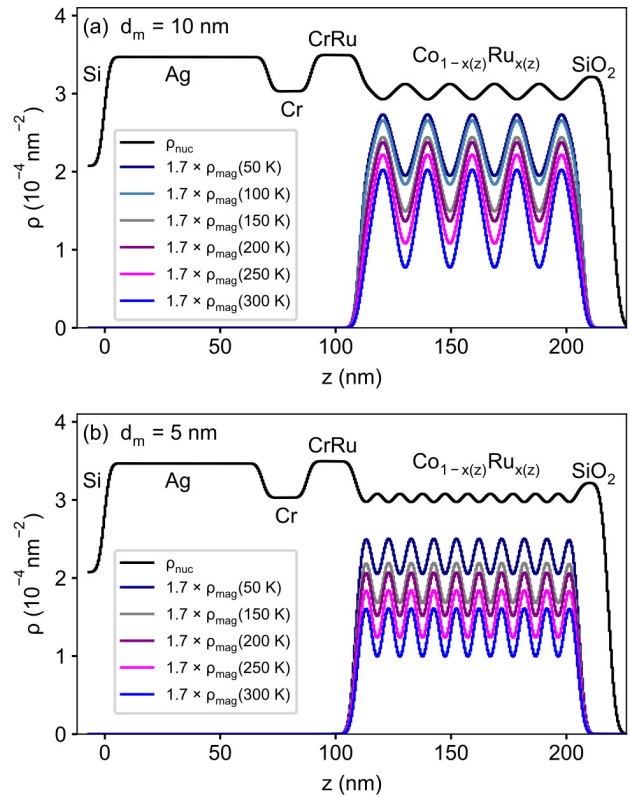}
\caption{\label{profiles}Scattering length density profiles determined for the (a) $d_{m}$ = 10 nm and (b) $d_{m}$ = 5 nm samples. The $T$-dependent $\rho_{mag}$ are scaled by a factor of 1.7 for easier visualization.}
\end{figure}

While qualitatively useful, a much more detailed, quantitative picture of the spatially dependent phase transition can be obtained by model-fitting the reflectivities over the full measured range \cite{majkrzak_polarized_2006}.   Figure \ref{RvQ} shows the fitted $T$-dependent $R^{++}$ and $R^{--}$ reflectivities for the (a) $d_{m}$ = 10 nm and (b) $d_{m}$ = 5 nm samples \footnote{The fitted data plotted on a larger scale is available in Section II of the online Supplemental Material}.  
In addition to the strong Bragg peaks, Fig. \ref{RvQ} shows pronounced oscillations at lower $Q$, indicating detailed sensitivity to the nuclear and magnetic depth profiles. At low $Q$ (i.e. away from the Bragg peaks),  the difference between $R^{++}$ and $R^{--}$ on average decreases with increasing $T$, indicative of the overall reduction in magnetization.  For each sample, data at all $T$ measured were simultaneously fit to a consistent scattering length density model with a $T$-independent nuclear profile.  Due to the complexity of the sample structure, we utilized highly constrained models, based on the nominal layer structure and composition.  With the exception of the SiO$_{2}$ cap (which could be degraded, etc.), values of $\rho_{nuc}$ were fixed to expected values for all layers.  This includes the Co$_{1-x(z)}$Ru$_{x(z)}$ alloy multilayers, which were modeled in terms of a repeating unit cell comprised of 20 sub-layers with $\rho_{nuc}$ corresponding to the nominal triangular $x$ profile.  The thicknesses of the layers and $d_{m}$ were treated as free parameters.  The $\rho_{mag}$ unit cell was also triangular and symmetric, and in registry with the nuclear unit cell.  It was parameterized in terms of the minima and maxima, with linearly varying $\rho_{mag}$ in between. Interlayer roughness $\sigma$ was accounted for in terms of an error function smoothing between layers, including between the very thin Co$_{1-x}$Ru$_{x}$ unit cell sublayers\footnote{Additional fitting parameters were employed to account for small variations in intensity normalization and sample alignment. These nuisance parameters were found to be effectively uncorrelated with parameters of interest.}.  For simplicity, $\sigma$ was assumed to be the same for each interface (again with the exception of the topmost cap interface), and was constrained to be the same for the nuclear and magnetic profiles.  Roughness was propagated across multiple interfaces, ensuring that the resulting profile corresponds to an appropriate convolution between roughness and the intended ``perfect'' triangular waveform\cite{maranville_distributed_2017}. The best-fit scattering length density profiles are shown in Figure \ref{profiles}, with selected model parameters are shown in Table \ref{params}.  
\begin{table}
\caption{\label{params}Selected best-fit parameters determined from PNR model fitting.}
\begin{ruledtabular}
\begin{tabular}{ccccc}
 nom. $d_{m}$ & meas. $d_{m}$ & $\sigma$  & $M^{min}$(50 K) & $M^{max}$(50 K) \\
 (nm) & (nm) & (nm) & (kA m$^{-1}$) & (kA m$^{-1}$)
 \\ \hline
 10 & 9.68 $\pm$ 0.01 & 1.97 $\pm${0.08}  & 357 $\pm$ 7 & 590 $\pm$ 7 \\
 5 & 4.90 $\pm$ 0.06 & 1.99 $\pm${0.06}   & 334 $\pm$ 12 & 586 $\pm$ 12 \\
\end{tabular}
\end{ruledtabular}
\end{table}

The nuclear profiles provide the dominant contribution to the scattering, and show features associated with the non-magnetic underlayers, and the oscillating Co$_{1-x(z)}$Ru$_{x(z)}$ multilayers.  As $\rho_{nuc}$ is larger for Ru than for Co, the high Ru $x$ = 0.31 regions are manifested as maxima in $\rho_{nuc}$, while minima correspond to $x$ = 0.21.  The magnetic scattering length density profiles of the Co$_{1-x(z)}$Ru$_{x(z)}$ are also highly modulated at all $T$.  While higher Ru concentration leads to higher $\rho_{nuc}$, it should of course lead to lower magnetization, and thereby lower $\rho_{mag}$ for high $x$.  We find this to be the case. Notably, we are also sensitive to the phase of the $\rho_{mag}$ oscillations, finding good fits only when $\rho_{mag}$($z$) is 180$^{\circ}$ out of phase with the corresponding $\rho_{nuc}$($z$), an affirmation of our choice of model.  As shown in Table 1, we find that the modulation thicknesses are within 3$\%$ of target values, and the interlayer roughness is essentially identical for the two samples.   Despite the heavily constrained modeling scheme, the PNR profiles result in excellent fits to the data, shown as solid lines in Figs. \ref{braggpeaks}-\ref{RvQ}.

\begin{figure*}
\includegraphics[scale=0.90,trim = 0cm 0cm 0cm 0cm,clip ]{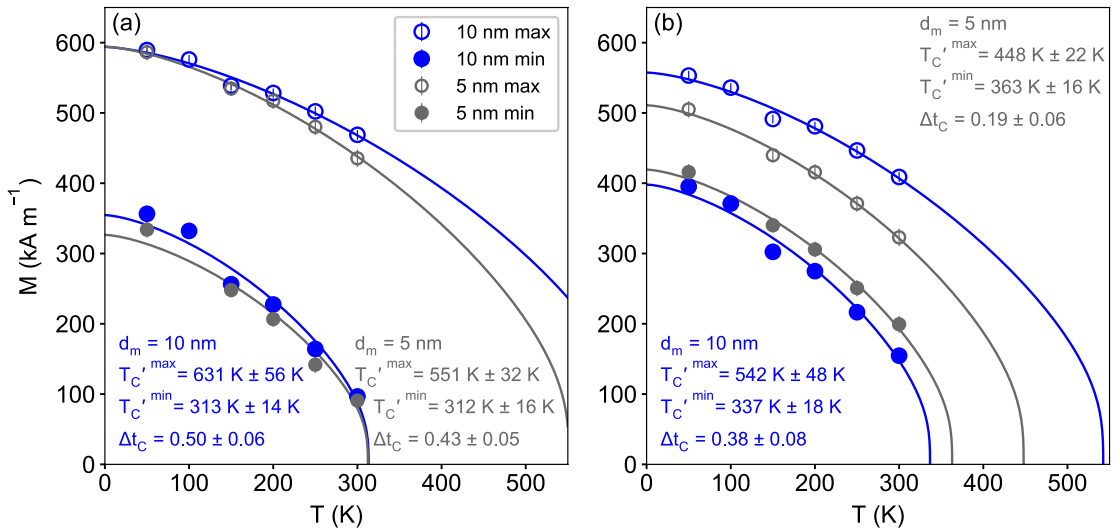}
\caption{\label{PNR_MvT}Temperature dependencies of the minimum (solid symbols) and maximum (hollow symbols) magnetizations for the d$_m$ = 10 nm and 5 nm Co$_{1-x(z)}$Ru$_{x(z)}$ films, as determined from PNR.  Two limiting case interpretations are considered:  (a) The magnetic depth profiles are perfect over an area $A{l}$,  and (b) the apparent profile smearing is indicative of the profiles for any local area smaller than $A_{n}$. Error bars for (b) are assumed to be the same as those explicitly calculated for the values in (a).}
\end{figure*} 

It is important to bear in mind that the reflectivities fundamentally provide sensitivity to the $\rho$($z$) profile used to calculate the reflectivities, and different choices of model parameterization could yield similar or identical profiles that lead to equally good fits to the data.  As such, care must be taken in interpreting the profiles and corresponding best-fit parameters.  Specifically, the nuclear profiles are significantly smeared compared to what would be expected from the designed composition profiles shown in Fig. \ref{xrd}(b), and while we have chosen to parameterize this in terms of a convolution of the intended profile with roughness, we cannot be certain of the lateral lengthscale over which the the apparent roughness is manifested.   

For these measurements, we expect that the nuclear and magnetic depth profiles correspond to an in-plane average over an area $A_{n}$ $\approx$ 70 $\mu$m$^{2}$ or less \cite{majkrzak_determination_2014,fallarino_magnetic_2017}$^{,}$\footnote{$A_{n}$ is determined from the coherent extent of the neutron wavepacket perpendicular to the direction of propagation.  In Ref. \onlinecite{majkrzak_determination_2014}, this value is shown to be $\lesssim$ 2 $\mu$m for our experimental setup, but recent unpublished measurements indicate it is closer to about 0.5 $\mu$m.  Here we have used the latter value to calculate the approximate maximum $A_{n}$ for our measurements.}.  As such, specular neutron reflectometry cannot determine the area over which inhomogeneities are present more precisely than $A_{n}$.  Therefore, it is possible that the profiles are less rough locally than they appear in Fig. \ref{profiles}, which would have consequences for our interpretation of the depth-dependent $M$($T$).  Although we cannot determine this degree of locality from the PNR data, we can bracket the plausible values of the $T$-dependent magnetization minima and maxima by considering two limiting case interpretations.

First, let us assume that over some local area $A_{l}$, the concentration profiles exhibit nearly perfect as-designed steps as shown in Fig. \ref{xrd}(b), and that the apparent roughness arises from long-range in-plane inhomogeneities (e.g. small layer thickness variations, substrate micro-roughness, etc.) that are manifested over an area larger than $A_{l}$ but less than $A_{n}$.  We then assume that the local magnetization profile tracks the locally perfect nuclear structure, and consider the underlying triangular wave magnetic profile with roughness deconvolved.  Within the bounds of this interpretation, the actual magnetization minima and maxima are simply the best-fit values of the $M^{max}$($T$) and $M^{min}$($T$) fitting parameters, which are shown in Figure \ref{PNR_MvT}(a).   For both samples, $M$($T$) differs markedly for the minima and maxima, with $T_{C}$$^{\prime}$ clearly lower for the minima.  Shown inset in Fig. \ref{PNR_MvT} are the values of $T_{C} \,^{\prime \, min}$ and $T_{C} \,^{\prime \, max}$ estimated using Eq. \ref{Kuzmin} with shape parameters fixed at $s$ = 2.9 and $p$ = 1.9, as described in Section II.  The resulting  $T_{C} \,^{\prime \, min}$ values are comparable to the Curie temperatures of the corresponding uniform reference samples shown in Fig.~\ref{SQUIDVSM_MvT}, indicating that effects of interlayer coupling become insignificant as the material passes its local $T_{C} \,^{\prime \, min}$.  It is convenient to define the normalized local Curie temperature modulation,
\begin{equation}
\Delta t_{C} = \frac{T_{C} \,^{\prime \, max}-T_{C} \,^{\prime \, min}}{T_{C} \,^{\prime \, max}},
\label{modulation}
\end{equation}
as a figure of merit for quantitatively describing variation in $T_{C}$$^{\prime}$.  We find that $\Delta t_{C}$ = 0.50 for $d_{m}$ = 10 nm, and $\Delta t_{C}$ = 0.43 for $d_{m}$ = 5 nm, comparable to, but less than the value of 0.59 corresponding to Curie temperatures of the uniform reference samples shown in Fig.~\ref{SQUIDVSM_MvT}.  

Alternatively, we can make the assumption that the smearing apparent in Fig. \ref{profiles} is representative of inhomogeneities down to length scales that will affect the magnetization.  In the bounds of this interpretation, the actual magnetizations correspond to the values shown in the Fig. \ref{profiles} (via conversion with Eq. \ref{rhoM}).  These curves are displayed in Figure \ref{PNR_MvT}(b), and show a $T_{C}$$^{\prime}$ modulation that is still quite pronounced for both samples, with $\Delta t_{C}$ = 0.38 for $d_{m}$ = 10 nm, and $\Delta t_{C}$ = 0.19 for $d_{m}$ = 5 nm.  Therefore, from these two limiting cases, we can estimate that 0.13 $<$ $\Delta t_{C}$ $<$ 0.56.  

\section{Model Calculations}
\begin{figure*}
\includegraphics[scale=.35,trim = 0cm 0cm 0cm 0cm,clip ]{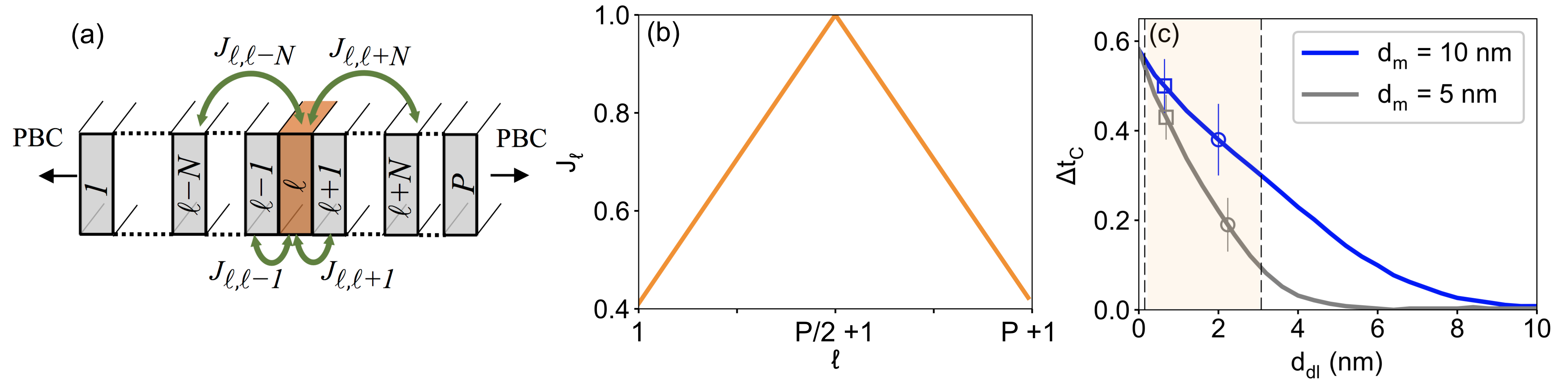}
\caption{\label{theory} (a) Depiction of the unit cell used for the MFA model.  $P$ atomic planes characterized by exchange strength $J_{\ell}$ are coupled to a variable number of neighbors $N$ on both sides with an exchange coupling strength $J_{\ell , \ell ^{\prime}}$. Periodic boundary conditions (PBC) are employed to mimic the sample multilayer structure.  (b) Exchange strength profile of the model unit cell.  (c) Local Curie temperature modulation as a function of exchange delocalization distance.  Solid lines correspond to the MFA model.  Open squares correspond experimental values derived from Fig. \ref{PNR_MvT}(a), while open circles correspond to values from Fig. \ref{PNR_MvT}(b).  Orange shading bracketed by dashed lines indicates range of uncertainty in $d_{dl}$ corresponding to 2 standard deviations.}
\end{figure*}

For the purpose of comparing our experimental results to theoretical expectations, we calculated the expected modulation of the magnetization for a model system that mimics the experimental one in the framework of the mean-field approximation (MFA) of the Ising model. In our previous work \cite{kirby_spatial_2016}, we considered only nearest neighbor exchange interactions to describe a gradient extending over 100 nm.  Here, we explore if our experimental data can be described theoretically and what level of exchange interaction confinement is compatible with the data. In turn, this means that within our model we must consider the possibility of exchange interactions that are not confined to only nearest neighbor spins or atoms.  Figure \ref{theory}(a) depicts the model unit cell used for calculations. 

The unit cell consists of layers $\ell$ (atomic planes) of thickness $d_{\ell}$ = 0.2 nm arranged along the $z$-axis. To relate to our samples, we constrain the number of layers $P = 2d_{m} d_{\ell} ^{-1}$.  Each spin within a given layer couples to spins in the same layer with an intrinsic exchange strength $J_{\ell}$, and to the spins in other layers $\ell ^{\prime}$ with an interlayer exchange coupling of strength $J_{\ell , \ell ^{\prime}}$ = $\sqrt{J_{\ell} \, J_{\ell ^{\prime}}}$.  Periodic boundary conditions (PBC) are employed on either side of the unit cell to mimic the multilayer structure of our samples.  The intrinsic exchange strength profile of the model is shown in Fig. \ref{theory}(b).  The model mimics the profile of the samples, with a triangular waveform of linearly varying $J_{\ell}$.  The minimum value of $J_{\ell}$ is set to 0.41, the ratio of the Curie temperatures of the two homogeneous $x$ = 0.31 and $x$ = 0.21 reference samples.  Using this exchange coupling profile, the ratio of the minimum and maximum local Curie temperatures matches the expected value from the reference samples exactly, assuming that each layer of the gradient material is isolated and no interlayer exchange coupling is present.

The magnetization of each layer is calculated self-consistently by solving the system of $P$ equations:
\begin{equation}
m_{\ell} \, = \, \textrm{tanh}\left[\frac{1}{T} h_{\ell}^{eff}\right], \, \ell \, = 1, \, ..., \, P.
\label{tanh}
\end{equation}
Here, $T$ is given in units normalized to the Curie temperature of the $x$ = 0.21 homogeneous sample.  The MFA effective field $h_{\ell}^{eff}$ arises because of the exchange coupling, and is defined as
\begin{equation}
\label{heff}
h_{\ell}^{eff} = \frac{1}{2N+1}\sum\limits_{\ell^{\prime} = \ell - N}^{\ell+N} J_{\ell,\ell^{\prime}}m_{\ell^{\prime}},
\label{sum}
\end{equation}
where $N$ is half the number of layers over which the exchange interaction is extended.  Alternatively, it is useful to define an exchange delocalization distance in absolute units,
\begin{equation}
\label{delocalization}
d_{dl} = 2Nd_{\ell}.
\label{sum}
\end{equation}
Particular cases are $N$ = 0, for which the layers are isolated, and $N$ = 1, for which we only have nearest-neighbor interactions.  Within this framework, we have calculated the $m_{\ell}$ magnetization profiles as functions of $T$ and $N$ for $d_{m}$ = 5 nm and $d_{m}$ = 10 nm.  By means of a subsequent analysis of these calculated profiles, we have determined the corresponding $T_{C} \,^{\prime \, min}$ (i.e. for the $\ell$ = 1, minimum $J$ layer), and $T_{C} \,^{\prime \, max}$ (i.e. for the $\ell$ = $\frac{P}{2}$ + 1, maximum $J$ layer), and thereby the corresponding values of the normalized local Curie temperature modulation, as functions of $d_{dl}$ (or $N$).  Fig. \ref{theory}(c) shows the MFA calculated $\Delta t_{C}$ for $d_{m}$ = 5 nm (gray lines) and $d_{m}$ = 10 nm (blue lines).   As the delocalization distance approaches zero, the individual layers become progressively more isolated, and $\Delta t_{C}$ for both values of $d_{m}$ converge to the value expected from the uniform reference samples.  As the delocalization distance increases, the layers become increasingly more coupled, the magnetization profiles become more homogeneous, and $\Delta t_{C}$ approaches zero.  

Experimental values of $\Delta t_{C}$ as determined from PNR are mapped on to the theoretical curves, depicted as open symbols in Fig. \ref{theory}(c).  Values corresponding to locally perfect depth profiles are shown as squares, with maximally rough profile values depicted as circles.  Notably, the experimental values for both the $d_{m}$ = 5 nm and the $d_{m}$ = 10 nm samples intersect the horizontal axis at nearly the same value of $d_{dl}$.  This similarity supports the appropriateness of our MFA model, as delocalization distance should be an intrinsic material parameter, independent of modulation distance.  Orange shading in (c) indicates the range of possible values of $d_{dl}$ based on our model and fitting uncertainty.  This shows that even under the most conservative assumptions, the delocalization distance is less than approximately 3 nm, and may be less than 1 nm if the samples are indeed more locally perfect than they appear to PNR. Delocalization over just one or even a few nm is somewhat counterintuitive, considering that the system studied is a metallic itinerant ferromagnet.  To put this range of $d_{dl}$ into perspective, we can compare to the magnetostatic exchange length $l_{ex}$, a commonly used parameter in micromagnetic calculations.  The exchange length is typically defined in terms of the relative strengths of the exchange and self-magnetostatic energies, and can be thought of as a length scale over which magnetic inhomogeneities are relevant in domain wall formation.  For pure Co, $l_{ex}$ $\approx$ 4 nm\cite{abo_definition_2013}, greater than even our largest estimate of $d_{dl}$.  
  
 \section{Conclusions}
We have shown that continuous composition gradients can be used to create continuous local Curie temperature gradients across nanometer lengthscales.  The observed temperature dependence of Bragg peaks in the PNR data directly demonstrate that $T_{C}$$^{\prime}$ is modulated over distances down to approximately 5 nm, while a combination of PNR model fitting and MFA calculations indicate localization distances of 3 nm or less. Despite the fact that ferromagnetism is essentially a collective effect, we find that the ferromagnetic phase transition evolves in a highly localized fashion, with depth-dependent interactions important only at distances less than a few nm.  The overall explanation for this localization is that for a layer at $T$ $>$ $T_{C}$$^{\prime}$ , the free energy costs to produce a spin polarized magnetic state is so high that it cannot be significantly magnetized by an adjacent layer at $T$ $<$ $T_{C}$$^{\prime}$  - even if the system is metallic and the magnetic order is mediated by itinerant electrons, as for the system described here.  For more insulating materials, the degree of magnetic property localization should only become more pronounced.  With this in mind, we assert that for virtually any such modulated ferromagnetic system, non-local materials properties are likely insignificant over length scales greater than about 3 nm, at which point the thermodynamic and corresponding behavior can be described purely in terms of local material properties.   Thus, nanoscale modulation of exchange strength, local Curie temperature, magnetization, and other magnetic properties should be achievable for a wide range of materials.  In addition to being fundamentally interesting, this degree of localization has important implications for magnetic devices and materials development \cite{sander_2017_2017}.  By recognizing the relative weakness of non-local effects, it is possible that novel collective effects can be achieved by utilizing simple nanoscale design rules.    

\begin{acknowledgments}
Work at nanoGUNE acknowledges support from the Basque Government under Project No. PI2015-1-19 and from the Spanish Ministry of Economy, Industry and Competitiveness under the Maria de Maeztu Units of Excellence Programme - MDM-2016-0618 and Project No. FIS2015-64519-R (MINECO/FEDER). Work at RIT was supported by NSF Award \#1609066.  P. R. acknowledges Obra Social la Caixa for her Ph.D. fellowship.  We thank Aaron Green of the University of Maryland for contributions to reflectometry modeling software, and Julie Borchers of NIST for helpful discussions.
\end{acknowledgments}

\end{document}